\documentclass[prl,reprint]{revtex4-1}
\usepackage{graphicx}
\usepackage{hyperref}
\usepackage{verbatim}
\usepackage{bbold}

\usepackage{amsmath}
\usepackage{amssymb}

\begin{document}

\title{Giant weak value amplification with chirped waveforms}
\author{Filippo M. Miatto$^{1}$}
\affiliation{$^1$Institute for Quantum Computing, University of Waterloo, Canada}
\date{\today}

\begin{abstract}
Weak value amplification is a classical phenomenon that can enhance the sensitivity of a measurement through clever use of interference. The most well-known paradigm of weak value amplification makes use of a Gaussian pulse, which is typical of pulsed laser systems. In this Letter we show that chirped pulses have a great advantage over Gaussians at detecting frequency shifts thanks to the large phase space area that they cover. As an example, we show that within the typical operative parameters of a radar, we can achieve two orders of magnitude amplification of small frequency shifts \emph{on top of the weak value amplification}. This idea could lead to new metrological avenues in the microwave optics domain, and to Doppler radar technology with unprecedented sensitivity.
\end{abstract}
\maketitle

\section{Introduction}
Weak value amplification (WVA) is an interference effect that can be exploited to magnify small perturbations \cite{PhysRevLett.60.1351}. In ideal conditions, it does not surpass the performance of optimized metrology \cite{PhysRevLett.113.120404}. However, such idealized measurements might just be too complicated to implement, or our measurement devices might present limitations \cite{arXiv:1402.0199,PhysRevX.4.011031, PhysRevLett.118.070802}. It is in these realistic situations that WVA can be of help. For a historical overview, and for an extensive survey of the literature on weak values, we refer to the review article: \cite{RevModPhys.86.307}. In this Letter we highlight the advantage that chirped pulses have over Gaussian pulses \cite{PhysRevLett.105.010405, PhysRevA.82.063822} at detecting frequency shifts within the WVA paradigm.

\section{Theory}

WVA requires three main ingredients \cite{RevModPhys.86.307}: an appropriate pair of physical systems, a weak interaction between them and a filtering process. The first system is typically named ``pointer'' as it acts as the hand on a dial. The second system, typically named ``selector'', is responsible for enabling the \emph{amplification} of the position of the hand on the dial. The weak interaction must couple these two systems in such a way that states of the pointer that differ by small amounts (i.e. that maintain a large overlap) are coupled to orthogonal states of the selector. Finally, the selector undergoes a filtering process. The role of filtering is to eliminate as much irrelevant signal as possible, leaving only the part which carries useful information. As we shall see, this protocol enhances the average displacement of the pointer, an effect known as ``weak value amplification''. 
To describe weak value amplification in a simple way we borrow the Dirac notation, but in no way we imply that this is a chiefly quantum effect.

We begin with selector and pointer in the pure uncorrelated states $|s_i,p_i\rangle$, where the subscript reminds us that these are the initial states. Such joint state then undergoes a weak unitary transformation $U(\epsilon)$, which by virtue of its weakness ($\epsilon\ll1$) can be approximated by $(1-i\epsilon \hat A\otimes \hat T)$, where $\hat A$ is a self-adjoint operator acting on the selector and $\hat T$ is the generator of transformations of the pointer. For instance, in the case of a birefringent crystal $\hat A$ could be $|H\rangle\langle H|-|V\rangle\langle V|$ (making the horizontal and vertical polarizations pick up opposite signs and therefore move in opposite directions at the same rate) and $\hat T$ would be the transverse momentum operator, which generates transverse displacements of a beam of light.

It is interesting and useful to rearrange the terms to highlight the ``weak value'', which is a complex number that regulates many of the aspects of these measurements. First, we apply the weak unitary to the initial state
\begin{align}
|s_i,p_i\rangle\rightarrow(1-i\epsilon \hat A\otimes \hat T)|s_i,p_i\rangle.
\end{align}
Then, we filter on a final selector state $|s_f\rangle$ and obtain the final (unnormalized) pointer state
\begin{align}
\label{unnormalized}
\biggl(\langle s_f|s_i\rangle-\langle s_f|\hat A|s_i\rangle\, i\epsilon\hat T\biggr)|p_i\rangle,
\end{align}
whose norm is the fraction of signal that makes it through the filtering process, which is $|\langle s_f|s_i\rangle|^2+O(\epsilon^2\langle T^2\rangle)$. The weak value appears by dividing Eq. \eqref{unnormalized} by $\langle s_f|s_i\rangle$ (our state is still unnormalized, as the superscript reminds us):
\begin{align}
|p_f^\mathrm{un}\rangle=\bigl(1-A_w\, i\epsilon\hat T\bigr)|p_i\rangle,
\end{align}
where $A_w=\langle s_f|\hat A|s_i\rangle/\langle s_f|s_i\rangle$ is the weak value of the operator $\hat A$, evaluated between the initial state $|s_i\rangle$ and the final state $|s_f\rangle$. Note that $A_w$ depends on both the initial and final state, which is of crucial, because we can control them and consequently we can control the real and imaginary parts of the weak value. At this point we re-introduce the appropriate normalization factor and end up with the final normalized pointer state
\begin{align}
|p_f\rangle=N\bigl(1-A_w\, i\epsilon\hat T\bigr)|p_i\rangle,
\end{align}
where $N=1/\sqrt{\langle  p_f^\mathrm{un}| p_f^\mathrm{un}\rangle}$ and 
\begin{align}
\langle  p_f^\mathrm{un}| p_f^\mathrm{un}\rangle=1+2\epsilon\langle T\rangle_i\mathrm{Im}(A_w)+\epsilon^2|A_w|^2\langle T^2\rangle_i
\end{align}
where $\langle T^n\rangle_i=\langle p_i|\hat T^n|p_i\rangle$ is the $n$-th moment of $\hat T$ evaluated on the initial state. 

We can now adopt one of two strategies, depending which quantity is easier to measure or which one performs better: \emph{i}) we measure the average value of the generator $\hat T$ or \emph{ii}) we measure the average value of the conjugate observable, i.e. the observable whose transformations are generated by $\hat T$. Typically (e.g. for a gaussian pointer state) in the first case the imaginary part of the weak value $A_w$ is responsible for amplification, while in the second case the real part is. A clever choice of initial and final states of the selector allows one to control which of the two parts of $A_w$ will have a large value. We will describe only the second strategy, which applies to our case.

The second strategy consists in measuring the observable that is conjugate to $\hat T$, which is the one transformed by the action of $U(\epsilon)$, i.e. the degree of freedom actually undergoing the weak unitary transformation. Let's call the observable corresponding to such quantity $\hat W$, we have that
\begin{align}
\label{averageW}
\langle W\rangle_f&=\langle p_f|\hat W|p_f\rangle=\int_{-\infty}^\infty w \tilde P_f(w)\,dw
\end{align}
where $\tilde P_f(w)=|\tilde p_f(w)|^2$ and $\tilde p_f(w)$ is the Fourier transform of $p_f(t)$:
\begin{align}
\label{FT}
\tilde p_f(w)=N\biggl(1+\epsilon A_w\frac{\partial}{\partial w}\biggr)\tilde p_i(w)
\end{align}
Again, assuming that $p_i(t)$ is even and centred on zero, we have that its Fourier transform is real, and by expanding $P_f(w)$ one obtains 
\begin{align}
\label{secondStrategy}
\langle W\rangle_f=N^2\bigg[\langle W\rangle_i&+2\epsilon\mathrm{Re}(A_w)\int_{-\infty}^\infty w\, \tilde p_i'(w)\tilde p_i(w) dw\nonumber\\
&+O(\epsilon^2)\bigg]
\end{align}
The expectation value of $\hat W$ is now complemented by a contribution that is amplified by the real part of the weak value.
However, note that it is not necessary to assume evenness of $p_i(t)$, as in the case of a chirped pulse. In those cases, one should start from Eq.~\eqref{FT} and apply it directly into Eq.~\eqref{averageW}.

By controlling the initial and final polarization states, one can subsequently amplify the signal. However, the weak value cannot be made large at will, as the weakness conditions must be preserved. In particular, we assumed that the interaction could be approximated to linear order, so one needs to be check that higher order terms can actually be neglected. In practice, the dependence of the average of $\hat T$ or $\hat W$ on the relevant component of the weak value must be linear. Additionally, one needs to consider that uncertainty over the initial and final states translates to an amplified uncertainty over $A_w$: if we were working with polarization as selector, we would have to consider small but relevant effects such as a non-zero extinction ratio and our accuracy and precision in rotating the polarizers. We perform a more detailed analysis in the next sections.

\section{Examples}
\subsection{Gaussian pulse}
In the first example we consider the initial pointer as a Gaussian pulse centered on frequency $\omega_0$ of duration $\tau$:
\begin{align}
p_i(t)=\frac{1}{(2\pi)^{1/4}\sqrt{\tau}}\exp\left(-\frac{t^2}{4\tau^2}\right)\exp(-i\omega_0 t)
\end{align}
The weak unitary interaction shifts the frequency $\omega_0$ by a small amount $\epsilon$, compatible with the weakness conditions.

Following the second strategy, we apply Eq.~\eqref{secondStrategy} and obtain
\begin{align}
\langle \omega\rangle_f\approx \omega_0-\epsilon \mathrm{Re}(A_w)+O(\epsilon^2)
\end{align}
which is in the linear regime for large enough bandwidths. Here the small frequency shift is amplified by a factor $\mathrm{Re}(A_w)$.

\subsection{Chirped pulse}
\begin{figure}[h]
\begin{center}
\includegraphics[width=\columnwidth]{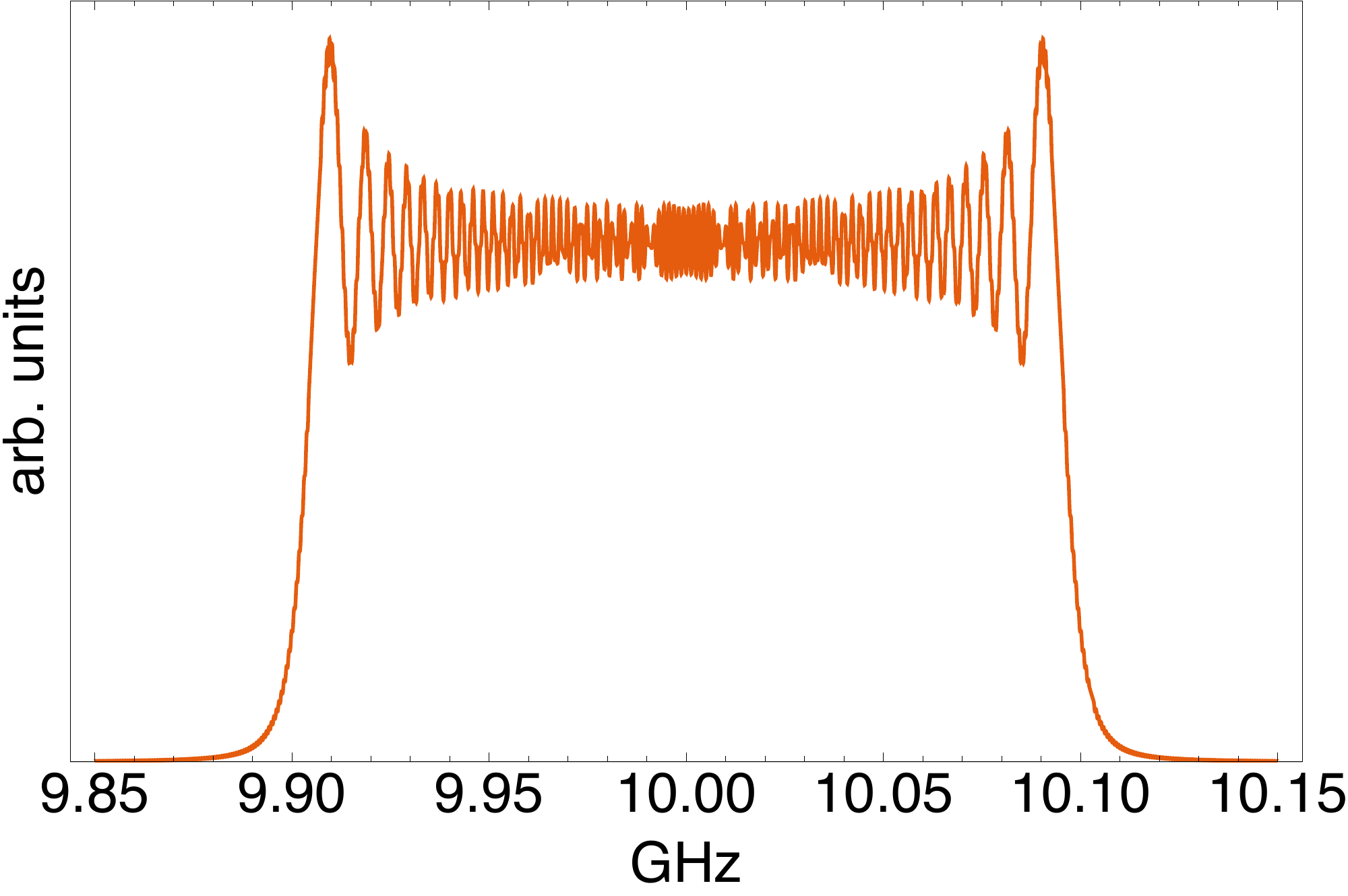}
\caption{\label{ChirpSpectrum}The power spectrum of the chirped pulse used in the example. It is centered at $\omega_0=10$ GHz and it sweeps a bandwidth of $200$ MHz in $10\ \mu$s.}
\end{center}
\end{figure}
In the second example, we consider a linearly chirped pulse. The pointer state is
\begin{align}
 p_i(t)=\frac{\mathrm{rect}(t/\tau)}{\sqrt{\tau}}\exp[i(\omega_0+R t)t]
\end{align}
where $R$ is the chirp rate, and where $\mathrm{rect}(x)$ is 1 for $-\frac12\leq x\leq \frac12$ and 0 otherwise. The Fourier spectrum of this function can be calculated analytically (see Fig.~\ref{ChirpSpectrum}):
\begin{align}
\label{actualSpectrum}
\tilde p_i(\omega)=\frac{e^{-i\frac{\Omega^2}{4R}}}{\sqrt{8\Delta}}\left[\mathrm{Erf}\left(\frac{\Delta+\Omega}{2\sqrt{i R}}\right)+\mathrm{Erf}\left(\frac{\Delta-\Omega}{2\sqrt{i R}}\right)\right]
\end{align}
where $\Omega=\omega-\omega_0$ and $\mathrm{Erf}(x)$ is the Gaussian error function. This expression is not too complicated to allow for precise numerical calculations. However, in order to carry out some first-order analytical calculations we approximate it with a rectangular spectrum centered in $\omega_0$ of width $2\Delta$:
\begin{align}
\tilde p_i(\omega)\approx \frac{1}{\sqrt{2\Delta}} \mathrm{rect}\left(\frac{\Omega}{2\Delta}\right)\exp\left[-\frac{i\Omega^2}{4R}\right].
\end{align}
Note that this approximation holds well in the linear regime, which is the regime of interest for WVA. 

By applying Eq.~\eqref{averageW} and Eq.~\eqref{FT} for a weak frequency shift $\epsilon$, we obtain
\begin{align}
\langle\omega\rangle_f\approx\omega_0+\epsilon\frac{\tau \Delta}{3}\mathrm{Im}(A_w)+O(\epsilon^2),
\end{align}
which holds as long as $\frac{\omega_0\epsilon}{4 R}|A_w|^2\ll\mathrm{Im}(A_w)$, otherwise we are departing from the linear regime. The formula above is telling us that a small frequency shift $\epsilon$ is amplified by the imaginary part of the weak value as well as by a factor proportional to the time-bandwidth product $\tau\Delta$.

How large can $\tau\Delta$ be? Chirps are not constrained by simple frequency-time Fourier relations as Gaussians are, because the frequency of a chirp is time-dependent. The time-bandwidth product for chirps is tunable and it can easily be in the order of hundreds for a typical radar chirp, whereas for Gaussians it is Fourier-limited to order 1: Gaussians actually would be a bad choice for this task because they \emph{minimize} the product of the quadratures! The value of non-Fourier limited waveforms was first pointed out in \cite{viza2013weak}.

To give a realistic example, we consider a chirp of duration $\tau\approx 10$ $\mu$s. We center it around $\omega_0=10$ GHz and we sweep a bandwidth of $2\Delta=200$ MHz \cite{skolnik1970radar}. This implies that the frequency must increase at a rate $R=10$ THz/s.
With these numbers we obtain $\langle\omega\rangle_f\approx\omega_0+333\epsilon\,\mathrm{Im}(A_w)$, which means that even for modest values of $\mathrm{Im}(A_w)$, the amplification would be hundreds of times larger than what we would obtain with WVA and a Gaussian pulse.

\section{On controlling the weak value}
\begin{figure}[h!]
\begin{center}
\includegraphics[width=\columnwidth]{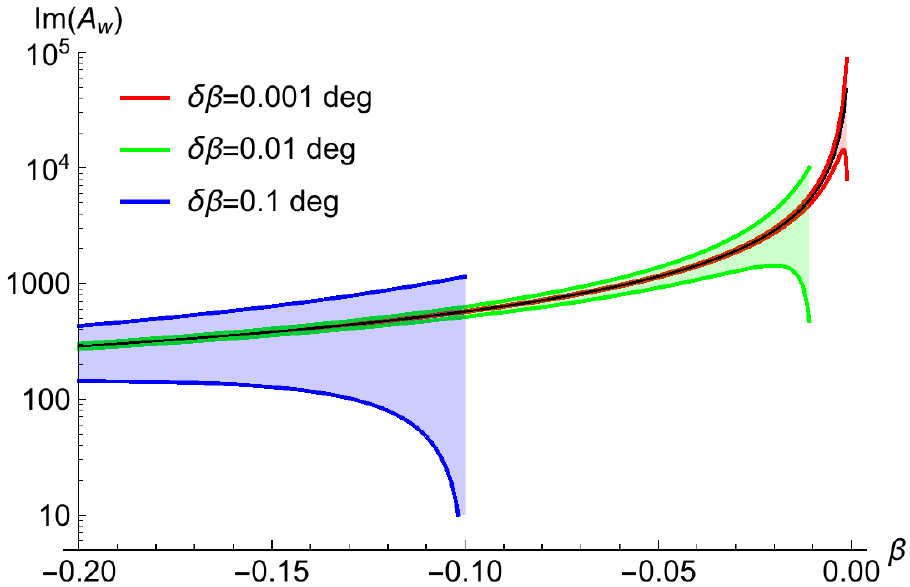}
\caption{\label{errors} If one desires to reach a large imaginary part of the weak value, one needs to control the polarization to a high degree of accuracy. In this figure we plot the weak value in black as a function of the ellipticity angle $\beta$ (in deg), and its $1\sigma$ error bands corresponding to three different ellipticity uncertainties $\delta\beta$.}
\end{center}
\end{figure}

\begin{figure}[h!]
\begin{center}
\includegraphics[width=\columnwidth]{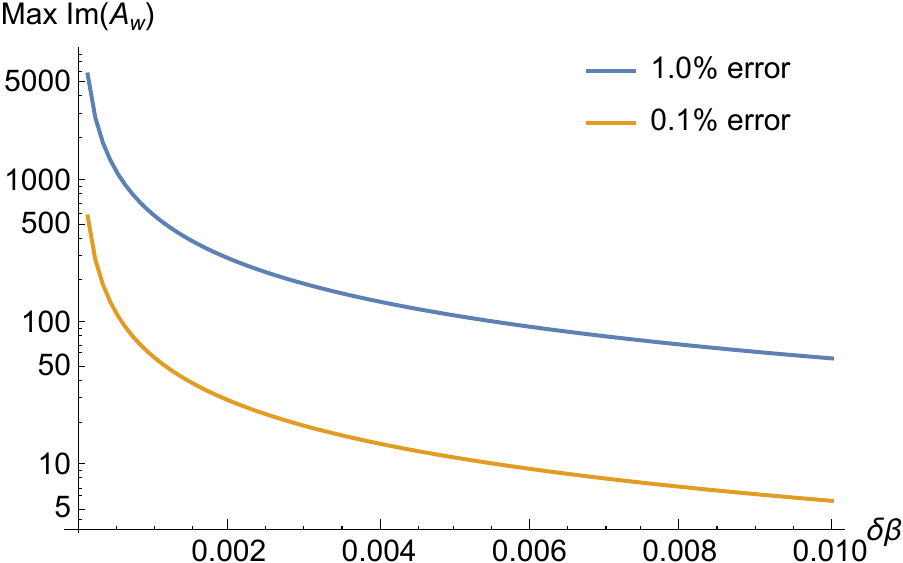}
\caption{\label{maxAwComplete} The largest value of $\mathrm{Im}(A_w)$ that we can achieve while tolerating a given error (here 1\% and 0.1\%) depends on the uncertainty on the experimental parameters. As one can see here, it is possible to attain $\mathrm{Im}(A_w)\approx 100\pm 1$ with an ellipticity uncertainty of 0.01 deg.}
\end{center}
\end{figure}

In this section we estimate a realistic upper bound for the weak value.
The weak value is controlled by creating the initial state $|s_i\rangle$, and by selecting the final state $|s_f\rangle$ with precision. The factors that influence the errors on $A_w$ are mainly the accuracy with which we can produce the initial state and the accuracy of the orientation of the final polarizer, and to a smaller extent also their extinction ratio and residual birefringence. To supply a meaningful example, we calculate the weak value between polarization states as follows: we choose
\begin{itemize}
\item an initial polarization $|s_i\rangle=\frac{1}{\sqrt{2}}(|H\rangle+e^{i\beta}|V\rangle)$ which is slightly elliptical ($\beta\ll1$), with equal contributions of H and V
\item an operator $\hat A=|H\rangle\langle H|$ (so that we have the interaction affect only the horizontal component).
\item a final linearly polarized state that is almost orthogonal to $|s_i\rangle$, i.e. $|s_f\rangle=\cos(\frac{\pi}{4}+\alpha)|H\rangle-\sin(\frac{\pi}{4}+\alpha)|V\rangle$
\end{itemize}
Obviously, disregarding extinction ratio (which for microwaves can be as low as 40 dB) and non-zero birefringence, there would be a singularity at $\alpha=\beta=0$, but the imperfections of a realistic polarizer prevent this divergence from happening. From simple propagation of uncertainty we learn that the absolute error on $\mathrm{Im}(A_w)$ eventually overcomes the weak value itself and that since we are interested in the imaginary part of $A_w$, the uncertainty in the ellipticity $\delta\beta$ matters much more than $\delta\alpha$. To illustrate this, let's evaluate the largest value of $\mathrm{Im}(A_w)$ that we can achieve while tolerating a 1\% error, in the case of an ellipticity uncertainty $\delta\beta$ of 0.1 deg, 0.01 deg and 0.001 deg. In Fig.~\ref{errors} we plot the weak value and the error that is due to the three different ellipticity uncertainties. Finally, in Fig.~\ref{maxAwComplete} to better understand how $\delta\beta$ affects the largest value of $\mathrm{Im}(A_w)$, we evaluate $\max[\mathrm{Im}(A_w)]$ under the constraint that the error be $1\%$ and $0.1\%$ for ellipticity uncertainties ranging from $\delta \beta=10^{-4}$ deg to $\delta \beta=10^{-2}$ deg.

So we can see that it is certainly possible to attain an imaginary weak value of the order of 100. If we compound it with a realistic time-bandwidth product also of order 100, we obtain an effective amplification factor of order $10^4$.

\section{Outlook and conclusion}
We have uncovered the value that chirps have over Gaussians for detecting frequency shifts within the WVA paradigm, but why stop here? Are chirps the best waveform for this task? What phase-space characteristics are necessary to allow for high sensitivity in general? Can we mimic what happens in this situation in other, very different contexts (e.g. for transverse position shifts)? What is the physical significance of the extra amplification parameter? We plan on investigating these questions in a future work.

In conclusion, this work lays the foundations for a new paradigm in microwave optical metrology, as well as for practical radar Doppler technology based on weak value amplification. At the same time, it gives an important example of the potential of non-standard waveforms to improve the sensitivity of the WVA technique.

\section{acknowledgements}
I thank Dr Carmine Clemente for helpful consultations regarding radar technology.

\bibliography{bibliography}
\bibstyle{unsrt}

\end{document}